\documentclass[a4paper,11pt]{article}
\pdfoutput=1 % if your are submitting a pdflatex (i.e. if you have
\usepackage{jcappub} % for details on the use of the package, please
\usepackage[T1]{fontenc} % if needed

\usepackage{bm}
\renewcommand{\v}[1]{\bm{#1} }
\newcommand{\vw}{{\mathrm{\textbf{\textsl{w}}}}}
\newcommand{\mw}{{\mathrm{\textsl{w}}}}

\newcommand{\sH}[0]{{\mathcal{H}}}
\newcommand{\vol}[2]{\hspace{-0.8mm}\mbox{$\text{d}^{\hspace{-0.0mm}#1}$}\hspace{-0.2mm}#2\hspace{0.8mm}\ }
\title{\boldmath Newton to Einstein -- dust to dust}

\author[a,b,c]{Michael Kopp,}
\author[a,c]{Cora Uhlemann,}
\author[a]{and Thomas Haugg}

\affiliation[a]{Arnold Sommerfeld Center for Theoretical Physics,\\  Ludwig-Maximilians University Munich, Theresienstr. 37, 80333 Munich, Germany}
\affiliation[b]{University Observatory,\\ Ludwig-Maximilians University Munich, Scheinerstr. 1, 81679 Munich, Germany}
\affiliation[c]{Excellence Cluster Universe, Boltzmannstr. 2, 85748 Garching, Germany}

\emailAdd{michael.kopp@physik.lmu.de}
\emailAdd{thomas.haugg@physik.lmu.de}
\emailAdd{cora.uhlemann@physik.lmu.de}

\abstract{We investigate the relation between the standard Newtonian equations for a pressureless fluid (dust) and the Einstein equations in a double expansion in small scales and small metric perturbations. We find that parts of the Einstein equations can be rewritten as a closed system of two coupled differential equations for the scalar and transverse vector metric perturbations in Poisson gauge. It is then shown that this system is equivalent to the Newtonian system of continuity and Euler equations. Brustein and Riotto (2011) conjectured the equivalence of these systems in the special case where vector perturbations were neglected. We show that this approach does not lead to the Euler equation but to a physically different one with large deviations already in the 1-loop power spectrum. We show that it is also possible to consistently set to zero the vector perturbations which strongly constrains the allowed initial conditions, in particular excluding Gaussian ones such that inclusion of vector perturbations is inevitable in the cosmological context. In addition we derive nonlinear equations for the gravitational slip and tensor perturbations, thereby extending Newtonian gravity of a dust fluid to account for nonlinear light propagation effects and dust-induced gravitational waves.}

\begin{document}
\maketitle
%\flushbottom
\section{Introduction}
\label{sec:intro}
In \cite{BR11}, a new method to study the evolution of nonlinear cosmological matter perturbations was presented in which the nonlinear Einstein equations were employed to deduce a single equation for the Newtonian potential. The key advantage of this approach is to provide a  closed and non-perturbative equation for the gravitational potential instead of a coupled fluid system for density and velocity. This allows to study directly the Newtonian potential which remains always small, even if density perturbations become large. The framework presented for the gravitational potential was shown to bear close resemblance to the Newtonian fluid formulation with regard to perturbative and mean field solutions and therefore their equivalence was conjectured. 

Following closely \citep{BR11}, we derive a coupled system for the Newtonian potential and a transverse vector field from the Einstein equations for a dust fluid by performing a small scale expansion and an expansion in the smallness of metric perturbations in Poisson gauge. Hereby taking vector perturbations $\omega_i$  of the metric explicitly into account we are able to prove the equivalence of parts of the Einstein equation and the Newtonian pressureless fluid equations. The remaining parts of the Einstein system yield nonlinear equations for  tensor perturbations $\chi_{ij}$ and the `slip' $\Psi - \Phi$ thereby naturally extending Newtonian gravity of dust to allow for a consistent description of Einsteinian effects like light propagation and gravitational waves. Similar to $\Psi$, the quantities $\Phi, \omega_i, \chi_{ij}$ are therefore already encoded in the Newtonian dynamics of a dust fluid and can be extracted from it. This fact was recently observed in \cite{BRW13},\footnote{We would like to thank Marco Bruni for making us aware of this work.} where $\omega_i$ was measured from a Newtonian N-body simulation. Although $\omega_i$ turned out to be sub-leading compared to $\Psi$ and therefore consistent with the double expansion scheme, it was on average 10 times larger than expected from a perturbative calculation \cite{BRW13}.

We will furthermore show that restricting attention to scalar metric perturbations leads to a constraint equation that amounts to considering  fine-tuned initial conditions. On the other hand, ignoring this constraint as done in \cite{BR11}  modifies the Euler equation. Both approaches for vanishing vector perturbations therefore have serious ramifications. Most notably, discarding the constraint on the initial conditions leads to disagreement with known standard perturbation theory results, which remained unacknowledged in \citep{BR11}. We re-derive perturbation theory including vector perturbations in Appendix \ref{App} and compare numerical results for the 1-loop matter and momentum power spectrum in Section \ref{sec:nullomega}. 
\section{Evolution equation in the presence of vector perturbations}
\label{sec:evolution}
We assume that the metric is perturbatively close to a flat Friedmann-Robertson-Walker (FRW) metric written in Poisson gauge and conformal time $\tau$ with scale factor $a(\tau)$
\begin{equation} \label{metric}
ds^2=a^2(\tau)\left[-e^{2\Phi}d\tau^2+2\omega_i d\tau dx^i+ \left(e^{-2\Psi}\delta_{ij}+\chi_{ij}\right)dx^{i}dx^{j}\right]\,,
\end{equation}
where $\Phi$ and $\Psi$ are assumed to be first order perturbation quantities in the initial conditions and at later times $\omega_i$ and $\chi_{ij}$ are secondarily induced, with $\omega_{i,i} = \chi_{ij,j} = \chi_{ii} = 0$. The physical justification for this is that throughout the universe, except very close to black holes and neutron stars, $\Phi$ and $\Psi$ are quasistatic and  remain at their primordially small value, typically $\Phi, \Psi = \mathcal{O}(\epsilon) \simeq 10^{-5}$. In addition any primordial $\omega_i$ and $\chi_{ij}$ will have decayed quickly such that we will assume that they vanish in the initial conditions and are only induced with size $\mathcal O(\epsilon^2$) later on. The vector perturbation $\omega_i$ grows slowly, and although $\chi_{ij}$ are constantly emitted they are weak and decay quickly.

Although the metric is perturbatively close to a FRW metric, spacetime curvature is not assumed to be perturbatively close to FRW on small scales. Spatial gradients $\nabla_i=\partial_i$ of metric perturbations, determining curvature, can become much larger. For example, in the quasilinear regime of structure formation, typically even if $\Psi \simeq 10^{-5}$ we have that $\Psi_{,i}/ \mathcal{H}\simeq v^i \simeq 10^{-3}$ and $\Delta \Psi /\mathcal{H}^2 \simeq \delta \simeq 1$, where $1/\mathcal{H}= a/a'$ is the comoving Hubble radius, $v^i$  matter velocity and $\delta$ the matter density contrast. We therefore introduce another small bookkeeping quantity $\eta$ and estimate spatial derivatives by assigning $\mathcal H^{-1}\partial_i = \mathcal O (\eta^{-1})$ and assuming that $\Delta \Psi /\mathcal{H}^2 = \mathcal{O}(\epsilon/\eta^2) \simeq \mathcal{O}(1)$. In Fourier space this means $k \gg \mathcal{H}$, and the expansion parameter $\eta$ is the ratio between the typical length scale of perturbations and the size of the Universe. We will see later in Eq.\,\eqref{CoupledMasterEqb} that the dynamical equations suggest that $\omega_i = \mathcal O(\epsilon^2/\eta)$ and therefore vector perturbations are a little bit more important than originally assumed. Recovering Newtonian gravity imposes this as a consistency requirement. In particular this implies that $\Delta \omega_i/\mathcal{H}^2 = \mathcal O(\epsilon^2/\eta^3)=\mathcal O(\epsilon/\eta)$ and $ \omega_{i,j}/ \mathcal{H} = \mathcal O(\epsilon)$.

Performing this expansion scheme on the Einstein tensor calculated from the perturbed FRW metric \eqref{metric}, see for example Eqs.\,(A.9)-(A.11) in \cite{BMR07}, one can easily recover the result obtained in \citep{BR11}, see Eqs.\,\eqref{gradExpG} below. In the $00$-component the leading order terms are $\mathcal{O}(1)$, the $0i$-component is $\mathcal O(\epsilon/\eta)$ and the $ij$-component order $\mathcal O(\epsilon)$. Therefore we assume a priori $\Delta(\Psi-\Phi)/\mathcal{H}^2=\mathcal{O}(\epsilon)$. This means that one can set $\Phi = \Psi$ everywhere except where the $\mathcal{O}(\eta^2 \epsilon)$ correction is not subleading, which happens only for $\Delta \Phi$ in the $ij$-component. Note also that for gravitational waves $\chi_{ij}$, time derivatives are as important as spatial derivatives, because they travel with the speed of light. Therefore we have a priori $\chi_{ij}''/ \mathcal{H}^2 = \mathcal{O}( \Delta \chi_{ij} / \mathcal{H}^2)= \mathcal{O}(\epsilon^2/\eta^2) =  \mathcal{O}(\epsilon)$. Taking into account all the aforementioned assumptions, that have to be checked a posteriori, the Einstein tensor takes the following form when keeping in each component only the leading orders in $\epsilon$ and $\eta$:
\begin{subequations} \label{gradExpG}
\begin{align}
G_{00}&=3\mathcal{H}^2+2\Delta\Psi\,, \\
G_{0i}&=2 \Psi_{,i}'+2\mathcal{H}\Psi_{,i} - \frac{1}{2} \Delta \omega_i\,, \label{0ieq}\\
G_{ij}&=\left[\left(\mathcal{H}^2-2\frac{a''}{a}\right)\left(1-4\Psi\right)+2\Psi''+6\mathcal{H}\Psi' +(\nabla\Psi)^2-\Delta\left(\Psi-\Phi\right)\right]\delta_{ij}-\\ \label{ijeq}
 \notag  &\quad  -2\Psi_{,i}\Psi_{,j}+\nabla_i\nabla_j\left(\Psi-\Phi\right) - \mathcal{H} (\omega_{i,j}+\omega_{j,i})- \frac{1}{2} (\omega_{i,j}'+\omega_{j,i}') + \\ \notag &\quad
 + \frac{1}{2} \left( \chi_{ij}'' - \Delta \chi_{ij}\right)\,. \notag
\end{align}
\end{subequations}
The following projectors 
\begin{align}
({\mathcal{P}_{\rm L}})^{i j} =  \frac{\nabla_i \nabla_j}{\Delta}\,, \qquad 
({\mathcal{P}_{\rm V}})^{ij}_k = \left( \delta^j_k - \frac{\nabla_k \nabla_j}{\Delta} \right) \nabla_i\,,
\end{align}
applied to  $G_{i j}=T_{i j}$ will be used in the following to derive closed equations of motion for the scalar $\Psi$ and vector $\omega_i$. We use units where $8\pi G =1$ and $c=1$. 
\paragraph*{Master equations}  Considering a dust fluid of density $\rho$ and four-velocity $u_\mu$ with energy momentum tensor $T_{\mu \nu} = \rho u_{\mu} u_{\nu}$ one can write its $ij$-component in terms of the $00$ and $0i$-components: $T_{ij}=T_{0i}T_{0j}/T_{00}$. One can then employ the Einstein equations $G_{\mu \nu}= T_{\mu \nu}$ to write a closed form equation for the metric 
\begin{equation} \label{dustEinsteinEq}
G_{ij}=\frac{G_{0i}G_{0j}}{G_{00}}\,,
\end{equation}
thus eliminating $\rho$ and $u_\mu$ from the equation.  The system of interest then consists of the longitudinal $({\mathcal{P}_{\rm L}})^{i j} $ and the vector $({\mathcal{P}_{\rm V}})^{ij}_k$ projections of Eq.\,\eqref{dustEinsteinEq} with Einstein tensor components \eqref{gradExpG}, in which $\Phi$ and $\chi_{ij}$ drop out automatically. Assuming that the Friedmann equations of an Einstein-de Sitter universe with average density $\bar \rho$ hold separately, the master system takes the form
\begin{subequations} \label{CoupledMasterEq}
\begin{align}
\label{CoupledMasterEqa}
\Psi'' +3\mathcal H\Psi' %-2\left(\mathcal{H}^2-2\frac{a''}{a}\right) \Psi
+\frac{1}{2}(\nabla\Psi)^2 &= ({\mathcal{P}_{\rm L}})^{i j} S_{ij}\,,\\ \label{CoupledMasterEqb}
\frac{1}{4}\Delta \omega_i'+\frac{1}{2}\Delta \omega_i \mathcal{H} &= ({\mathcal{P}_{\rm V}})^{km}_i  S_{km}\,,
\end{align}
where we defined the source tensor $S_{ij}:= \Psi_{,i}\Psi_{,j} + \tfrac{1}{2} G_{0i}G_{0j}/G_{00}$, whose explicit form is
\begin{align} \label{source}
S_{ij} &= \left(  \Psi_{,i}\Psi_{,j}+\frac{2}{3\mathcal{H}^2}\left\{  \frac{\left[\left(\Psi'+\mathcal{H}\Psi \right)_{,i}- \frac{1}{4}\Delta \omega_i\right] \left[\left(\Psi'+\mathcal{H}\Psi\right)_{,j}- \frac{1}{4}\Delta \omega_j\right]}{1+\frac{2}{3\mathcal{H}^2}\Delta\Psi }\right\}\right)\,.
\end{align}
\end{subequations}
Contrary to what one might naively expect, vector perturbations are crucial in order to recover Newtonian gravity \cite{BRW13}. We will prove this for the case of a pressureless fluid in Section \ref{sec:equiv}.

The master system \eqref{CoupledMasterEq} does not contain all information present in Eq.\,\eqref{dustEinsteinEq}. The remaining bits can be extracted similarly by applying
\begin{subequations}
\begin{align}
({\mathcal{P}_{\rm TT}})_{ij}^{km} &=  \left( \delta_{im} - \frac{\nabla_i \nabla_m}{\Delta} \right)\left( \delta_{jk} - \frac{\nabla_j \nabla_k}{\Delta} \right)- \frac{1}{2} \left( \delta_{km} - \frac{\nabla_k \nabla_m}{\Delta} \right) \left( \delta_{ij} - \frac{\nabla_i \nabla_j}{\Delta} \right) \,,\\ 
({\mathcal{P}_{\rm TL}})^{i j} &=  \delta_{ij} - 3 \frac{\nabla_i \nabla_j}{\Delta}
\end{align}
\end{subequations}
to Eq.\,\eqref{dustEinsteinEq}. The resulting equations determine $\chi_{ij}$ and $\Phi$ as
\begin{subequations} \label{Rest}
\begin{align}
\label{Resta}
\Delta(\Psi - \Phi)
 & = ({\mathcal{P}_{\rm TL}})^{i j} S_{ij}\,,\\ \label{Restb}
 \frac{1}{4}( \chi''_{ij} -\Delta \chi_{ij})  & = ({\mathcal{P}_{\rm TT}})^{km}_{i j}  S_{km}\,.
\end{align}
\end{subequations}
Since $\chi_{ij}$ and $\Phi$ do not influence the dynamics of $\Psi$ and $\omega_i$, we do not consider these equations in the following. The equations \eqref{Rest} can be applied to calculate nonlinear light propagation effects, like estimating nonlinear corrections to gravitational lensing and the Sachs-Wolfe effect, or the gravitational waves induced by nonlinear structure formation. All metric perturbations $\Psi,\Phi$, $\omega_i$ and $\chi_{ij}$ are in general generated by nonlinearities and necessary to calculate light propagation, see \cite{RMKB96} for perturbative treatment. In 
\cite{BRW13} the effect of $\v{\omega}$ on weak gravitational lensing was estimated from $\v{\omega}$ determined via Eq.\,\eqref{jdefconsequence} by measuring $\v{\nabla} \times [(1+\delta)\v{v}]$ in a N-body simulation.

Finally let us note that it is straightforward to include a cosmological constant by simply replacing  $G_{00} \rightarrow G_{00}-\Lambda a^2$ and $G_{ij} \rightarrow G_{ij}+\Lambda a^2 (1+2 \Psi) \delta_{ij} $ in Eq.\,\eqref{gradExpG}, see App.\,\ref{sec:masterwithpress}.

\section{Equivalence of fluid and Einstein systems} \label{sec:equiv}
In \citep{BR11} the possible equivalence of \eqref{CoupledMasterEqa} with $\v{\omega}=0$  and the nonlinear Newtonian pressureless fluid equations was mentioned. However, this issue has not been investigated further nor been resolved in a conclusive manner. The goal of this section is to show that the set of Newtonian fluid equations is indeed equivalent to the Einsteinian Eqs.\,\eqref{CoupledMasterEq}. As we will point out in the next section, the constraint arising from forcing $\v{\omega}\equiv0$, which was not taken into account in \citep{BR11}, is incompatible with general initial conditions. In this section we therefore keep the transverse vector $\v{\omega}$ unconstrained, apart from the original assumption that $\v{\omega}= \mathcal{O}(\epsilon^2/\eta)$.  
\paragraph*{Fluid equations} Introducing the momentum $\v{j}= (1+\delta) \v{v}$, the curl-free non-relativistic fluid equations (which also follow from $\nabla^\mu G_{\mu \nu} = \nabla^\mu (\rho u_\mu u_\nu)$, with $G_{\mu \nu}$ from Eq.\,\eqref{gradExpG},\footnote{Note that the Bianchi identity $\nabla^\mu G_{\mu \nu} =0$ does not hold anymore for the $G_{\mu \nu}$ with components \eqref{gradExpG} and $\nabla^\mu$, the covariant derivative within the $\epsilon,\eta$ expansion scheme, see Eqs.\,(A.8) of \cite{BMR07}. However $\nabla^\mu G_{\mu \nu} = \nabla^\mu T_{\mu \nu}$, consistently expanded in $\epsilon$ and $\eta$ leads to the correct Newtonian equations (\ref{jequations}).} $\rho=3\mathcal H^2(1+\delta)$, $u_0=-a$ and $u_i = a v^i$ and the aforementioned assumptions) can be written as
\begin{subequations} \label{jequations} 
\begin{align}
\delta '+\v{\nabla}\cdot \v{j}&= 0\,, \label{jConti}\\
\v{j}'+\mathcal{H} \v{j}  +\v{\nabla}\cdot\left(\frac{\v{j} \v{j}}{1+\delta} \right)+ (1+\delta) \v{\nabla}\Psi&= 0\,, \label{jEuler}\\
\v{\nabla}\times \v{v} &=0\,. \label{jCurl}
\end{align}
\end{subequations}
The Poisson equation supplements both the fluid equations and the master system
\begin{align}
\delta &= \frac{2}{3 \mathcal{H}^2} \Delta \Psi\,. \label{jPoiss}
\end{align}
\newpage
\paragraph*{Equivalence}
\paragraph*{\eqref{CoupledMasterEq} $\Rightarrow$ \eqref{jequations}}
The Euler equation \eqref{jEuler} can be derived easily from the master equation by taking the time derivative of 
\begin{align} 
\label{jdef}
\v{j} := -\frac{2}{3 \mathcal{H}^2} \left[ \v{\nabla} (\Psi' + \mathcal{H} \Psi)-\frac{1}{4} \Delta \v{\omega} \right]
\end{align}
and replacing $\Psi''$ and $\Delta \v{\omega}'$ according to \eqref{CoupledMasterEqa} and \eqref{CoupledMasterEqb}, respectively. For details see App.\,\ref{sec:details1}. The continuity equation \eqref{jConti} is obtained by taking the time derivative of the Poisson equation and making use of the definition of $\v{j}$, Eq.\,\eqref{jdef}. Note that \eqref{jequations} is obtained with built-in condition $\v{\nabla}\times \v{v} =0$. This is because $\v{\omega}$ in the Einstein system \eqref{CoupledMasterEq} is a second order quantity and therefore initially $\v{v} = \v{j}$. But the fluid equations then imply $\v{\nabla}\times \v{v} =0$ for all later times since the Euler equation \eqref{jEuler} for $\vw := \v{\nabla} \times \v{v}$ implies
\begin{equation}
\vw'+\mathcal{H} \vw - \v{\nabla}\times(\v{v}\times \vw) =0\,,
\end{equation}
and guarantees that if $\vw=0$ initially, it remains so. Note that the initial condition $\vw=0$ does not constrain $\delta$ and $\theta := \v{\nabla}\cdot \v{v}$. It is also important to note that $\vw =0$ does not imply $\v{\omega} =0$ since 
\begin{align} 
\label{wandomega}
%\vw = -\frac{2}{3 \mathcal{H}^2}\v{\nabla} \times \frac{ \left[ \v{\nabla} (\Psi' + \mathcal{H} \Psi)-\frac{1}{4} \Delta \v{\omega} \right]}{1+ \frac{2}{3 \mathcal{H}^2} \Delta \Psi}\,.
\vw = \frac{\v{\nabla}\times\Delta \v{\omega}}{6 \mathcal{H}^2 (1+\delta)} - \frac{\v{\nabla} \delta}{(1+\delta)^2} \times \v{j}\,.
\end{align}

\paragraph*{\eqref{jequations} $\Rightarrow$ \eqref{CoupledMasterEq}} To derive the coupled master system \eqref{CoupledMasterEq} from the fluid equations \eqref{jequations}, one has to define a transverse vector $\v{\omega}$ according to \eqref{jdef}. The longitudinal $\nabla_i/\Delta$ and transverse part $(\delta_{ij} - \nabla_i  \nabla_j/\Delta)$ of the Euler equation \eqref{jEuler} can then be used to derive \eqref{CoupledMasterEqa} and \eqref{CoupledMasterEqb}, respectively with the help of \eqref{jPoiss} and \eqref{jConti}. For details see App.\,\ref{sec:details}. 
\paragraph*{Remarks} We would like to point out that the equivalence might break down after shell-crossing infinities $\delta \rightarrow \infty$ occur in the fluid system \eqref{jequations}. These infinities are an artifact of the assumed single streaming  pressureless fluid and seem to be harmless in the master system \eqref{CoupledMasterEq} since they simply correspond to regions where the matter source term (the curly brackets in \eqref{source}) vanishes. This vanishing happens only if the numerator of the matter source remains finite at shell crossings, which might not be the case.  It would be interesting to extend the framework of \cite{BR11} to include multi-streaming effects, which would allow to describe dark matter dynamics at even smaller scales and times after shell crossings. However, simply adding a shear term to the energy momentum tensor $T_{\mu \nu} \rightarrow \rho u_\mu u_\nu +\sigma_{\mu \nu}$ requires either an additional dynamical equation for $\sigma_{\mu \nu}$ or to postulate $\sigma_{\mu \nu}$ to be a functional of $G_{00}$ and $G_{0i}$. This goes beyond the scope of this paper. In App.\,\ref{sec:masterwithpress} we show how to derive the corresponding master equation in the special case where $\sigma_{\mu \nu} = p(T_{00},T_{0i}) (u_{\mu} u_\nu + g_{\mu \nu})$ and outline the case of shear viscosity.

One might wonder why Poisson gauge equipped with the assumptions about the metric and its derivatives only, leads exactly to the Newtonian limit. The reason is that these assumptions imply for the Einstein tensor $G_{ij} \ll G_{0i} \ll G_{00}$ and therefore via Einstein equations  $T_{ij} \ll T_{0i} \ll T_{00}$, which together with small metric perturbations $\chi_{ij} , \omega_i \ll \Psi = \Phi \ll 1$ defines a Newtonian source. Therefore the $\eta$-$\epsilon$-expansion seems to be equivalent to the post-Friedmann expansion proposed in \cite{BRW13}.

It should be also noted that the recently described \cite{R13} cosmological frame dragging effect on dust disappears in the double expansion scheme used here. Although a nonzero $\v{\omega}$ is generated in our case, it leaves the dynamics of dust unchanged from the Newtonian case. $\v{\omega}$ encodes the miss-alignment of directions of the Newtonian $\v{\nabla} \delta$ and $\v{v}$, see Eq.\,\eqref{wandomega}. This was already mentioned in \cite{BRW13} such that considering vanishing vector perturbations $\v{\omega}=0$ enforces the constraint $\v{\nabla} \times \rho \v{v} = 0$, which was described in \cite{BRW13} as unphysical.

It is also interesting to note that the Newtonian system \eqref{jequations} manifestly contains only two scalar degrees of freedom $\delta$ and $\theta$, because $\vw = 0$ is a constant of motion.  The fact that if $\v{\omega}=0$, \eqref{CoupledMasterEqa}  contains only the two scalar degrees of freedom $\Psi$ and $\Psi'$  is suggestive for considering the case of $\v{\omega}=0$, which will be done in the next section.

\section{\boldmath Problems with $\omega_i =0$} \label{sec:nullomega}
The special case of vanishing vector perturbations (setting $\v{\omega}\equiv 0$) leads to the following system of differential equations
\begin{subequations} \label{CoupledRMasterEq}
\begin{align}
\label{CoupledRMasterEqa}
\Psi'' +3\mathcal H\Psi' %-2\left(\mathcal{H}^2-2\frac{a''}{a}\right) \Psi
+\frac{1}{2}(\nabla\Psi)^2 &= ({\mathcal{P}_{\rm L}})^{i j} S_{ij}[\v{\omega} =0]\,,\\ \label{CoupledRMasterEqb}
0 &= ({\mathcal{P}_{\rm V}})^{km}_i  S_{km}[\v{\omega} =0]\,,
\end{align}
\end{subequations}
with $S_{ij}[\v{\omega}\!=\!0]$ from Eq.\,\eqref{source}.
Translating the system \eqref{CoupledRMasterEq} into the fluid language, we observe that Eqs.\,\eqref{jequations} are still implied but that in addition to $\v{\nabla} \times \v{v} = 0$ also $\v{\nabla} \times \v{j} = 0$ is enforced during time evolution. This puts strong constraints on the initial conditions of $\delta$ and $\v{v}$. Constraint \eqref{CoupledRMasterEqb} is equivalent to the requirement that $\v{\nabla}\delta$ is aligned with the velocity $\v{v}$,
\begin{equation}
\v{\nabla} \delta \times \v{v} =0 \,, \label{Omegaconstraint}
\end{equation}
see Eq.\,\eqref{wandomega}.
While ignoring the constraint \eqref{CoupledRMasterEqb} is inconsistent, keeping the constraint has unwanted physical consequences for the allowed initial conditions of $\Psi$. In \cite{BR11}, only Eq.\,\eqref{CoupledRMasterEqa} without the accompanying constraint was obtained due to a mistake in going from Eq.\,(2.11) to Eq.\,(2.12) in \cite{BR11}.\footnote{While the left hand sides of the $0i$ and $ij$ Einstein equations, Eqs.\,(2.5) and (2.6) in \cite{BR11}, were projected onto their longitudinal parts, the right hand sides were not projected. The wrong Eq.\,(2.5) was then used in Eq.\,(2.11), leading to the master equation Eq.\,(2.13), or our Eq.\,\eqref{CoupledRMasterEqa} without the constraint \eqref{CoupledRMasterEqb}.} 
The master equation Eq.\,\eqref{CoupledRMasterEqa} considered in \cite{BR11} is equivalent to the following non-perturbative fluid-like system of equations
\begin{subequations} \label{Rjequations} 
\begin{align}
\delta '+\v{\nabla}\cdot \v{j}&= 0\,, \label{RjConti}\\
\v{\nabla}\cdot\left(\v{j}'+\mathcal{H} \v{j}  +\v{\nabla}\cdot\left(\frac{\v{j} \v{j}}{1+\delta} \right)+ (1+\delta) \v{\nabla}\Psi \right)&= 0\,, \label{RjEuler}\\
\v{\nabla}\times \v{j} &=0\,, \label{RjCurl}
\end{align}
\end{subequations} 
see  App.\,\ref{sec:details2}. These equations are only equivalent to the fluid equations \eqref{jequations} if $\v{\nabla} \delta \times \v{v} =0$. If this constraint is not satisfied, then although $\v{\nabla} \times \v{v}=0$ holds initially it is not conserved during time evolution.
\begin{center}
\begin{figure}[t]
\includegraphics[width=0.47\textwidth]{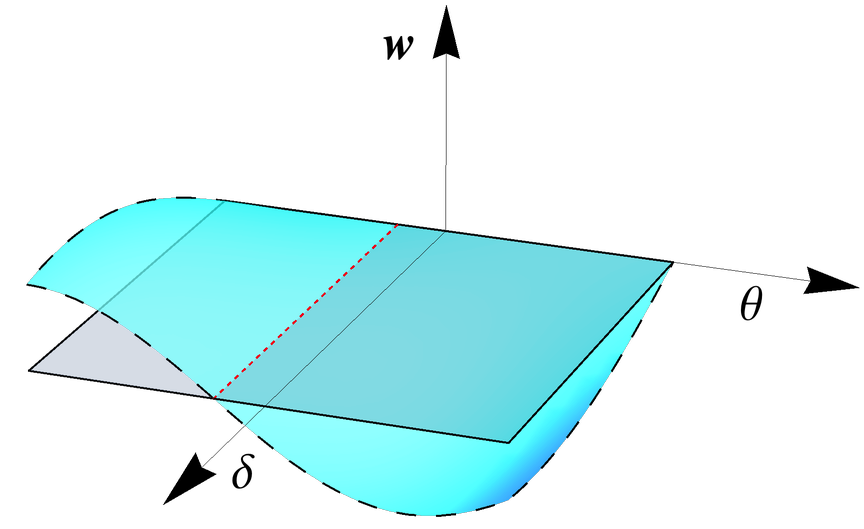} \qquad
\includegraphics[width=0.47\textwidth]{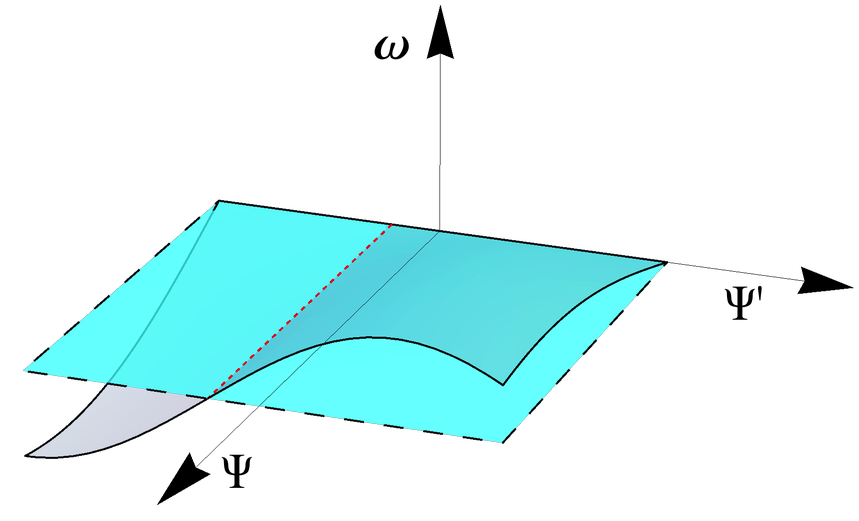}
\caption{A sketch of the configuration space of two scalars and one transverse vector, \textit{left}: in the basis $\delta,\theta,\vw$, \textit{right}: in the basis $\Psi,\Psi',\v{\omega}$.  The solution space with initial conditions $\vw=\v{\omega}=0$ is indicated by the two surfaces. The surface with the continuous boundary line corresponds to the solution space of Newtonian fluid equations \eqref{jequations} or \eqref{CoupledMasterEq} for which $\vw=0$ holds during time evolution. The surface with dashed boundary corresponds to the solution space of \eqref{Rjequations} or \eqref{CoupledRMasterEqa} for which  $\v{\omega}=0$ holds during time evolution, discarding \eqref{Omegaconstraint} or \eqref{CoupledRMasterEqb}, respectively. The dotted line -- the intersection of the two planes --  corresponds to the solution space of \eqref{CoupledRMasterEq} or (\ref{Omegaconstraint},\ref{Rjequations}) in which $\vw=\v{\omega}=0$ holds during time evolution. For instance, spherically symmetric solutions lie in this subspace.}
\label{fig:CompSolutions}
\end{figure}
\end{center}
 As we will see next, the solution to Eqs.\,\eqref{Rjequations} are not a good approximation to the solution of the Newtonian fluid system Eqs.\,\eqref{jequations} in perturbation theory. Hence, results obtained from \eqref{CoupledRMasterEqa}, corresponding to Eq.\,(2.13) from \citep{BR11}, should be reconsidered carefully using the full master system \eqref{CoupledRMasterEq}.
% It is beyond the scope of this paper to investigate whether results of this mathematically inconsistent\footnote{Inconsistent from the point of view of our assumptions in expanding the Einstein equations and subsequent derivation of the master equation. Of course there is nothing inconsistent or wrong with taking  Eq.\,(2.13) of \citep{BR11} or the fluid-like Eqs.\,\eqref{Rjequations} as a starting point.} approach are in good agreement with Newtonian N-body simulations.
We summarize the three  approaches of describing nonlinear dust dynamics in Fig.\,\ref{fig:CompSolutions}. 
\paragraph{Perturbation theory} If one wants to solve the $\v{\omega}=\vw=0$ system Eq. \eqref{CoupledRMasterEq} up to order $n$ in perturbation theory, one is forced to fine-tune $\delta_1,..,\delta_{n-1}$ and $\theta_1,...,\theta_{n-1}$, such that the constraint is satisfied to order $n$. In particular, it can be easily shown  that constraint \eqref{Omegaconstraint} is incompatible with Gaussian initial conditions for $\Psi_1$ determining $\delta_1\propto\Delta\Psi_1$ and $\v{v}_1\propto \v{\nabla}\Psi_1$. To this end we show that the expectation value  of $| \v{\nabla}\delta_1\times\v{v}_1|$ is nonzero:
\begin{subequations}
\begin{align}
& 0\neq \langle \left| \v{\nabla}\delta_1\times\v{v}_1\right| \rangle  \\
\quad \Leftrightarrow \quad & 0\neq  \langle (\v{\nabla}\delta_1)^2(\v{v}_1)^2-(\v{\nabla}\delta_1\cdot \v{v}_1)^2\rangle \\
\quad \Leftrightarrow \quad & 0\neq  \left(\int_0^\infty k^4P_1(k)\ dk\right)\left( \int_0^\infty P_1(p)\ dp\right) - \left(\int_0^\infty k^2P_1(k) dk \right)^2 \,,
\end{align}
\end{subequations}
where we used the definition of the linear power spectrum $\langle \delta_1(\v{k})  \delta_1(\v{p})\rangle = (2 \pi)^3 \delta_{\rm D}(\v{k} + \v{p}) P_1(k)$ and Wick's theorem.\\
On the other hand if one considers the $\v{\omega}=0$ system Eq.\,\eqref{CoupledRMasterEqa} as was done in \cite{BR11}, one obtains a wrong result for the perturbation theory kernels $F_{n\geq 3}$ defined as 
\begin{equation}
\delta_n(\v{p}) = \int \frac{d^3 k_1}{(2\pi)^3 } ... \frac{d^3 k_n}{(2\pi)^3 } (2\pi)^3 \delta_{\rm D}(\v{k}_1+...+\v{k}_n - \v{p} ) F_n(\v{k}_1,...,\v{k}_n) \delta_1(\bm{k}_1)... \delta_1(\bm{k}_n)\,.
\end{equation}
Indeed contrary to what was claimed in \cite{BR11}, the symmetrized perturbation kernel $F_3$ calculated from the master equation \eqref{CoupledRMasterEqa}, see Eqs.\,(3.19-20) in \cite{BR11} and our Appendix \ref{App}, is not equivalent to the symmetrized kernel $F_3$ found in the standard literature, see for example (A3) in \cite{GGRW86}. 
Since this kernel directly affects $P_{13,\delta\delta}$, defined by $2 \langle \delta_1(\v{k})  \delta_3(\v{p})\rangle = (2 \pi)^3 \delta_{\rm D}(\v{k} + \v{p}) P_{13,\delta\delta}(k)$, this leads to a discrepancy with results obtained for  one-loop power spectrum $P_{\delta\delta}=P_{1}+P_{22, \delta \delta}+P_{13, \delta\delta}$, using standard perturbation theory \cite{MSS91} based on Eqs.\,\eqref{jequations} and Gaussian statistics
\begin{center}
\begin{figure}[t]
\includegraphics[width=0.5\textwidth]{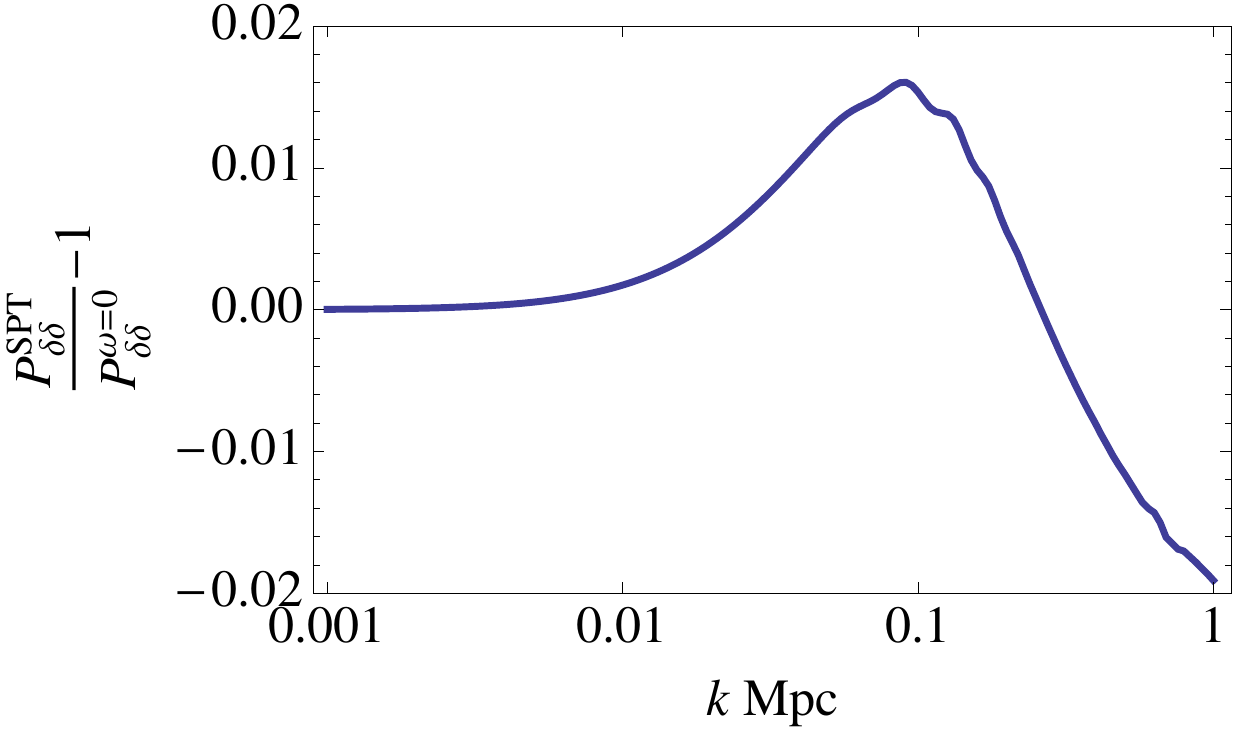}
\includegraphics[width=0.48\textwidth]{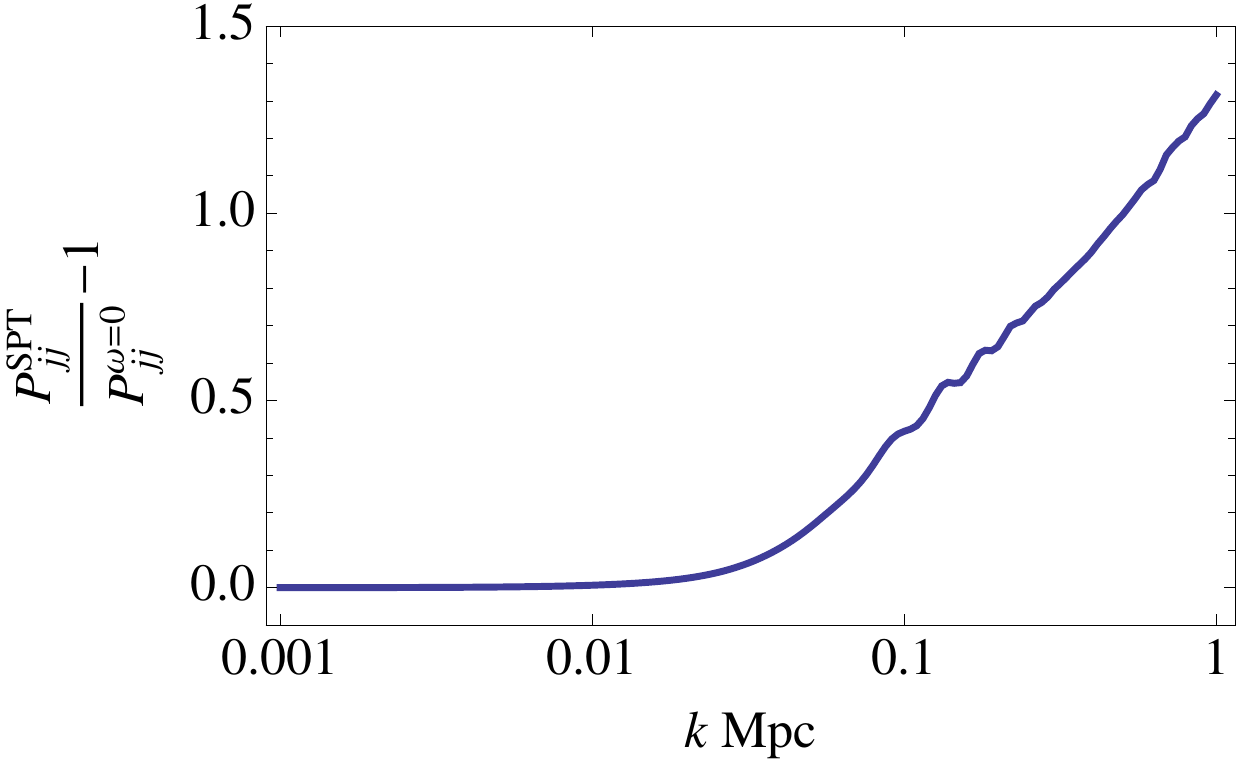}\\
\caption{\textit{Left}:  Comparison between one-loop matter power spectra $P_{\delta \delta}$   obtained using \eqref{P13R} and \eqref{P13SPT}, respectively.  \textit{Right}: Comparison between one-loop momentum power spectra $P_{jj}$ obtained using \eqref{Pjj}. By SPT we denote results obtained from the master system \eqref{CoupledMasterEq} including a growing mode for $\v{\omega}$, which we checked to be equivalent to SPT results based on \eqref{jequations}. The $\omega=0$ labeled power spectra are obtained from \eqref{CoupledRMasterEqa}, neglecting \eqref{CoupledRMasterEqb}.}
\label{fig:CompR}
\end{figure}
\end{center}
\begin{subequations}
\label{P13}
\begin{align}
P_{13,\delta\delta}^{\omega=0}(k) &= \frac{k^3}{252\cdot 4\pi^2} P_1(k) \int_0^\infty dr\ P_1(kr)\Bigg\{ \frac{40}{r^2}-\frac{614}{3}+\frac{440}{3}r^2-70r^4 +\label{P13R}\\
\notag &\qquad\qquad\qquad\qquad\qquad\qquad\qquad+\frac{5}{r^3}(r^2-1)^3(7r^2+4) \ln\left|\frac{1+r}{1-r}\right| \Bigg\}\,, \\
P_{13,\delta\delta}^{\text{SPT}}(k) &= \frac{k^3}{252\cdot 4\pi^2} P_1(k) \int_0^\infty dr\ P_1(kr) \Bigg\{\frac{12}{r^2}-158+100r^2-42r^4 +\label{P13SPT}\\
\notag &\qquad\qquad\qquad\qquad\qquad\qquad\qquad+\frac{3}{r^3}(r^2-1)^3(7r^2+2)\ln\left|\frac{1+r}{1-r}\right| \Bigg\}\,.    
\end{align}
\end{subequations}
If we had used statistics that guarantee Eq.\,\eqref{CoupledRMasterEqb}, both expressions would be identical.\footnote{See also \cite{RR13,BHMW13} for another situation in which non-Gaussian initial conditions are required in order to fulfill a relativistic constraint.} 
Numerical results obtained on the basis of \eqref{P13} indicate that the differences for $P_{13,\delta \delta}$ are of order unity and relevant only on large scales, which results in a quite small percent-level deviation on scales below 50 Mpc for the 1-loop matter power spectrum $P_{\delta\delta}=P_{1}+P_{22, \delta \delta}+P_{13, \delta\delta}$ with standard values of the cosmological parameters, see Fig.\,\ref{fig:CompR}, left.\footnote{The expression for $P_{22,\delta \delta}$ is the same in the two cases. See App.\,\ref{sec:2order}.} One might therefore hope that ignoring the constraint arising from $\v{\omega} = 0$, although inconsistent leads to physically acceptable results also in the non-perturbative regime for which Eq.\,\eqref{CoupledRMasterEqb} was devised in \cite{BR11}.  Unfortunately, the fact that the matter power spectrum in perturbation theory is close to SPT seems to be accidental, because, as we will show next, the momentum power spectrum shows much larger deviations. We therefore must  conclude that the master equation Eq.\,\eqref{CoupledRMasterEqb} cannot serve as an approximation to the full system Eqs.\,\eqref{CoupledMasterEq} in the nonlinear regime.

To see this, we first note that setting $\v{\omega}= 0$ gives rise to vorticity $\v{\nabla}\times\v{v}_2 \neq 0$ in second order perturbation theory, unless one takes the constraint \eqref{CoupledRMasterEqb}, at second order into account. From Eq.\,\eqref{jdef}, see also Eq.\,\eqref{v2sol}:
\begin{equation} \label{omegacon2nd}
v^i_2 \supset \Delta \Psi_1 \Psi_{1,i} \stackrel{\eqref{CoupledRMasterEqb}}{=}\frac{\nabla^i \nabla^j}{\Delta} (\Delta \Psi_1 \Psi_{1,j})\,.
\end{equation}
Therefore assuming $\v{\omega}= 0$ and discarding the constraint $\v{\nabla} \delta \times \v{v} =0$ introduces not only deviations in the matter power spectrum at third order but also a curl in the velocity field in second order perturbation theory. This curl has a large impact on the momentum power spectrum, see Fig.\,\ref{fig:CompR}, right.
The momentum $\v{j}$, \eqref{jdef} can be rewritten with the help of the Poisson equation as
\begin{equation} \label{jdefconsequence}
\v{j} = -\frac{\v{\nabla} \delta'}{\Delta} + \frac{1}{6\sH^2} \Delta \v{\omega}.
\end{equation}
The momentum power spectrum is defined via $\langle \v{j}(\v{k})\cdot \v{j}(\v{p}) \rangle= (2\pi)^3 \delta_{\rm D}(\v{k} + \v{p})  P_{jj}(k)$ and takes the following form in second order perturbation theory
\begin{subequations}
\label{Pjj}
\begin{align}
P_{jj}^{\omega=0}(k) &= \frac{\sH^2}{k^2}\left(P_{1}(k) + 4 P_{22,\delta\delta}(k) + 3 P_{13,\delta\delta}^{\omega=0}(k)\right)\\
P_{jj}^{\text{SPT}}(k) &=   \frac{\sH^2}{k^2}\left(P_{1}(k) + 4 P_{22,\delta\delta}(k) + 3 P_{13,\delta\delta}^{\rm SPT}(k)\right)+ P_{22,\omega\omega}^{\rm SPT}(k)\\
P_{22,\omega\omega}^{\rm SPT}(k) &= \frac{\sH^2 k}{2\cdot 4\pi^2} \int_0^\infty dr \int_{-1}^1 dx\ P_1(k\sqrt{1-2r x+r^2}) P_1(kr) \frac{(1-x^2)(1-2 rx)^2}{(1-2r x+r^2)^2}\,. 
\end{align}
\end{subequations}
Note that we are focusing here on momentum instead of the usual velocity power spectrum, because the 1-loop velocity power spectrum in case of $\v{\omega}=0$ suffers from a UV divergence, see App.\,\ref{sec:thirdorder}. This divergence is another hint that the system of fluid-like equations \eqref{Rjequations} is unphysical.

\section{Conclusion}
The double expansion in small scales ($\eta$) and the small potentials ($\epsilon$) of the Einstein equations with a dust fluid contains the Newtonian fluid equations \eqref{jequations} in form of the master equations \eqref{CoupledMasterEq}, if one assumes that metric vector perturbations are present and of order $\v{\omega} = \mathcal{O}(\epsilon^2/\eta)$. Additionally, this scheme also predicts the $\mathcal{O}(\epsilon)$ quantities $\Delta(\Psi -\Phi)$, and $\Delta \chi_{ij}$, corresponding to the slip and tensor perturbations. Although they are not relevant for the dynamics of the dust fluid itself, they extend Newtonian gravity to consistently include effects of light propagation, like nonlinear contributions to gravitational lensing or the Sachs-Wolfe effect, and gravitational waves. Closely related to this work is \cite{BRW13}, which is based on a post-Friedmann expansion of the small metric perturbations in powers of $c^{-1}$ in Poisson gauge, which for the dust case considered here seems to be equivalent to the $\epsilon$-$\eta$-expansion. In \cite{BRW13} the vector $\v{\omega}$ was measured in a Newtonian N-body simulation and the effect on the weak lensing convergence power spectrum was estimated. With the same methods also $\Delta(\Psi -\Phi)$ and $\Delta \chi_{ij}$ could be obtained and their effect on lensing estimated.

In the context of linear relativistic perturbation theory -- linear in $\epsilon$, non-perturbative in $\eta$ -- \cite{FS12, HHK12} found that the relativistic dynamics of dust can be mapped to Newtonian dynamics and therefore all relativistic information is encoded and can be extract from Newtonian simulations (at the linear level).

We showed that only inclusion of vector perturbations makes the master  \eqref{CoupledMasterEq}  and fluid systems \eqref{jequations} with standard initial conditions equivalent. Forcing $\v{\omega}\equiv0$, significantly truncates the allowed space of initial conditions. In particular inflationary initial conditions where the gravitational potential $\Psi$ is a Gaussian random field are excluded. Ignoring this constraint in \eqref{CoupledRMasterEq}, as was effectively done in \cite{BR11}, results in non-standard  fluid-like equations \eqref{Rjequations} and deviations in perturbation theory that become manifest at second order for velocity and third order for the density perturbations.  The 1-loop momentum power spectrum shows 50\% deviations at 10\,Mpc scales compared to SPT, while the 1-loop velocity power spectrum is not even converging, suggesting the unphysical nature of the master equation studied in \cite{BR11}.

The coupled nonlinear equations of motion for metric perturbations $\v{\omega}$ and $\Psi$, Eq.\,\eqref{CoupledMasterEq}, should be used as a starting point for investigations following the route of \cite{BR11} where Eq. \eqref{CoupledRMasterEqa} was used. We established that our result shares the remarkable feature found by \cite{BR11} that the matter source term, the curly brackets in  \eqref{source}, are switched off once the density contrast $\delta$ becomes large. 
Therefore the master system \eqref{CoupledMasterEq} for $\Psi$ and $\v{\omega}$ will prove useful in understanding the quasistatic dynamics  and decay of the Newtonian potential $\Psi$ in nonlinear structure formation.

\acknowledgments
We would like to thank Ram Brustein, Stefan Hofmann, Antonio Riotto, Cornelius Rampf, Dennis Schimmel and Jochen Weller for enlightening discussions and comments on the draft.  The work of MK \& CU was supported by the DFG cluster of excellence `Origin and Structure of the Universe'. The work of TH was supported by TR33 `The Dark Universe'.

%\paragraph{Note added.} This is also a good position for notes added after the paper has been written.

\newcommand{\apj}{Astrophys. J.}
\newcommand{\prd}{Phys. Rev. D}
\newcommand{\physrep}{Phys. Rep.}
\newcommand{\jcap}{J. Cosmol. Astropart. Phys.}
\bibliographystyle{plain}
\bibliography{CorrectedRbib}

\appendix 
\section*{Appendix}

\section{Explicit calculation of the equivalence } \label{sec:details}
\subsection{ between \eqref{jequations} and \eqref{CoupledMasterEq} } \label{sec:details1}
The derivative of $\v{j}= -2/(3 \sH^2)[ \v{\nabla}(\Psi' + \sH \Psi)-\Delta\v{\omega}/4]$ is given by
\begin{align} \label{jtimeder}
{j^i}' &= \sH j^i-\frac{2}{3 \sH^2}\left[(\Psi''+\sH \Psi' - \frac{1}{2}\sH^2 \Psi)_{,i} - \frac{1}{4} \Delta \omega_i' \right]
\end{align}
We now write the master equation in terms of $\v{j}$ and $\delta =2/(3\sH^2)\, \Delta \Psi$ and take the gradient of the $\Psi$ equation \eqref{CoupledMasterEqa}
\begin{subequations} \label{CoupledMasterEqApp}
\begin{align}
(\Psi'' +3\mathcal H\Psi' +\frac{1}{2}(\nabla\Psi)^2)_{,i} &= \frac{\nabla_k\nabla_m}{\nabla^2} \nabla_i \left(  \Psi_{,k}\Psi_{,m}+\frac{3 \sH^2}{2} \frac{j^m j^k}{1+\delta}\right) \label{CoupledMasterEqAppa}\\
\frac{1}{4}\Delta \omega_i'+\frac{1}{2}\Delta \omega_i \mathcal{H} & = \left(\frac{\nabla_i \nabla_m}{\Delta}- \delta^m_i  \right) \nabla_k \left(  \Psi_{,k} \Psi_{,m} + \frac{3\mathcal{H}^2}{2}  \frac{j^k j^m}{1+\delta}\right)
\end{align}
\end{subequations}
and subtract the second from the first equation
\begin{align} \label{step3App}
(\Psi'' +3\mathcal H\Psi' +\frac{1}{2}(\nabla\Psi)^2)_{,i} -\frac{1}{4}\Delta \omega_i'-\frac{1}{2}\Delta \omega_i \mathcal{H} & =  \left(  \Psi_{,k} \Psi_{,i} + \frac{3\mathcal{H}^2}{2}  \frac{j^k j^i}{1+\delta}\right)_{,k}
\end{align}
We now have an expression for the square bracket in Eq.\,\eqref{jtimeder}
\begin{align} \notag
\left[(\Psi''+\sH \Psi' - \frac{1}{2}\sH^2 \Psi)_{,i} - \frac{1}{4} \Delta \omega_i' \right] &= -2 \sH \Psi_{,i}' - \frac{1}{2} \sH^2 \Psi_{,i}+\Delta \Psi  \Psi_{,i}+\frac{1}{2} \sH \Delta \omega_i + \frac{3\mathcal{H}^2}{2} \left( \frac{j^k j^i}{1+\delta}\right)_{,k}\\
&= \frac{3 \sH^2}{2} \left( 2 \sH j^i + (1+\delta) \Psi_{,i}+  \left( \frac{j^k j^i}{1+\delta}\right)_{,k}\right)\,,\label{step2App}
\end{align}
where we used in the second line the Poisson equation and the definition of $\v{j}$. Equation \eqref{jtimeder} then becomes
\begin{align} \label{jEulerApp}
{j^i}' &= - \sH j^i - (1+\delta) \Psi_{,i}-  \left( \frac{j^k j^i}{1+\delta}\right)_{,k}\,,
\end{align}
which is just the Euler equation \eqref{jEuler}. Of course it is only the Euler equation (usually written in terms of $\v{v} = \v{j}/(1+\delta)$) if also the continuity equation $\delta' = - \v{\nabla}\cdot \v{j} $ holds.
In order to show the other direction we can simply follow all the steps backwards. We start with \eqref{jEulerApp} and insert the definition of $\v{\omega}$, Eq.\,\eqref{jdef}, which from the Newtonian point of view is just the curl part of $\v{j}$, while the longitudinal part is fixed by the continuity and Poisson equations, see also Eq.\,\eqref{jdefconsequence}. Therefore we can derive Eq.\,\eqref{step2App} with this definition of $\v{\omega}$. After reversing the algebraic manipulations from \eqref{step2App} to \eqref{step3App}, we only need to project onto the longitudinal and transverse parts to obtain the master system \eqref{CoupledMasterEqApp}, which is equivalent to \eqref{CoupledMasterEqa} if we assume vanishing boundary conditions.
\subsection{ between \eqref{Rjequations} and \eqref{CoupledRMasterEqa} } \label{sec:details2}
The derivative of $\v{j}= -2/(3 \sH^2) \v{\nabla}(\Psi' + \sH \Psi)$ is given by
\begin{align} \label{jRtimeder}
{j^i}' &= \sH j^i-\frac{2}{3 \sH^2}\left[(\Psi''+\sH \Psi' - \frac{1}{2}\sH^2 \Psi)_{,i} \right]
\end{align}
We now write the master equation in terms of $\v{j}$ and $\delta$ and take the gradient of the $\Psi$ equation \eqref{CoupledRMasterEqa}
\label{CoupledRMasterEqApp}
\begin{align}
(\Psi'' +3\mathcal H\Psi' +\frac{1}{2}(\nabla\Psi)^2)_{,i} &= \frac{\nabla_k\nabla_m}{\nabla^2} \nabla_i \left(  \Psi_{,k}\Psi_{,m}+\frac{3 \sH^2}{2} \frac{j^k j^m}{1+\delta}\right)\,, \label{CoupledRMasterEqAppa}
\end{align}
which can be used to eliminate $\Psi''$ in Eq.\,\eqref{jRtimeder}. Using the definition of $\v{j}$, we get
\begin{align} \label{jREulerAppb}
{j^i}' &=  -\sH j^i - \Psi_{,i} - \frac{2}{3\sH^2} \nabla_i \left( -\frac{1}{2} (\nabla \Psi)^2+  \frac{\nabla_k \nabla_m}{\Delta} \left(\Psi_{,k} \Psi_{,m} + \frac{3 \sH^2}{2}\frac{j^k j^m}{1+\delta}\right) \right)\,.
\end{align}
Using again the definition of $\v{j}$ and the Poisson equation this can be simplified to
\begin{align} \label{jREulerAppb}
{j^i}' &= - \sH j^i - \frac{\nabla_i \nabla_j}{\Delta}\left[ (1+\delta) \Psi_{,j}+  \left( \frac{j^k j^j}{1+\delta}\right)_{,k} \right]\,,
\end{align}
which is equivalent to 
\begin{subequations} \label{RjequationsApp} 
\begin{align}
\v{\nabla}\cdot\left(\v{j}'+\mathcal{H} \v{j}  +\v{\nabla}\cdot\left(\frac{\v{j} \v{j}}{1+\delta} \right)+ (1+\delta) \v{\nabla}\Psi \right)&= 0\,, \label{RjEulerApp}\\
\v{\nabla}\times \v{j} &=0\,. \label{RjCurlApp}
\end{align}
\end{subequations} 
Since all steps can be reversed we have shown the equivalence between the master equation of \cite{BR11}, our  \eqref{CoupledRMasterEqa} and the fluid like equation \eqref{Rjequations}.
\section{\boldmath Perturbation theory}\label{App}
We follow \citep{BR11} to expand the master system Eq.\,\eqref{CoupledMasterEq}
\begin{subequations}  \label{MasterEq}
\begin{align} \label{MasterEqPsi}
&\Psi'' +3\mathcal H\Psi' %-2\left(\mathcal{H}^2-2\frac{a''}{a}\right) \Psi
\\\quad&= -\frac{1}{2}(\nabla\Psi)^2 +\frac{\nabla^i\nabla^j}{\Delta} \left(  \Psi_{,i}\Psi_{,j}+\frac{2}{3\mathcal{H}^2}\left\{  \frac{\left[\left(\Psi'+\mathcal{H}\Psi \right)_{,i}- \tfrac{1}{4} \Delta \omega_i\right] \left[\left(\Psi'+\mathcal{H}\Psi\right)_{,j}- \tfrac{1}{4} \Delta \omega_j\right]}{1+\frac{2}{3\mathcal{H}^2}\Delta\Psi }\right\}\right)\,, \notag \\ 
&\frac{1}{4}\Delta \omega_i'+\frac{1}{2}\Delta \omega_i \mathcal{H} \label{MasterEqOmeg} \\ \quad & = \left(\frac{\nabla^i \nabla^m}{\Delta}- \delta^m_i  \right) \nabla^k \left(  \Psi_{,k} \Psi_{,m} + \frac{2}{3\mathcal{H}^2} \left\{  \frac{\left[\left(\Psi'+\mathcal{H}\Psi \right)_{,k}- \tfrac{1}{4} \Delta \omega_k\right] \left[\left(\Psi'+\mathcal{H}\Psi\right)_{,m}- \tfrac{1}{4} \Delta \omega_m\right]}{1+\frac{2}{3\mathcal{H}^2}\Delta\Psi }\right\} \right) \notag
\end{align}
\end{subequations} 
perturbatively as in Eqs.\,(3.1)-(3.20) of \citep{BR11}.  For convenience, this is done for the case of matter domination. We do this in order make it easier for the reader of \citep{BR11} to understand where and how perturbative solutions presented in \citep{BR11} are modified through the inclusion of $\v{\omega}$. The underscored equation numbers below correspond to equations in \citep{BR11}. Since we have already proven the nonperturbative equivalence of the master system \eqref{MasterEq} and the standard Newtonian dust fluid system \eqref{jequations}, it does not come as a surprise that a perturbative expansion of \eqref{MasterEq} agrees with standard perturbation theory (SPT), based on \eqref{jequations}, see for instance \cite{GGRW86}. The Friedmann equations and background quantities read
\begin{align}
\label{Friedmann}
3 \sH^2=a^2 \bar \rho , \quad \left(\sH^2-2\frac{a''}{a}\right)=0, \quad \bar \rho= \bar \rho_0\frac{a_0^3}{a^3} \quad \Rightarrow \quad a=\tau^2, \quad \sH=\frac{2}{\tau}\,.
\end{align} 
$\Psi$ and $\v{\omega}$ are expanded up to third order employing that $\v{\omega}$ is a second order quantity
\begin{align} \label{pertexp}
\Psi &= \Psi_1 + \Psi_2  + \Psi_3 + ...\,, \\ 
\omega_i &=\phantom{ \Psi_1+}  \omega_i^{(2)}  +\omega_i^{(3)} + ...\,.\notag
\end{align}
The standard perturbative expansion does not treat gradients in a special way. Therefore we have to remember that although we estimated in the non-perturbative case, for instance, that $\Delta \v{\omega} = \mathcal{O}(\epsilon^2/\eta^3)=  \mathcal{O}(\epsilon/\eta)$ and $\Delta \Psi /\sH^2 =  \mathcal{O}(\epsilon/\eta^2) \simeq 1$, we have now $\Delta \v{\omega} = \mathcal{O}(\epsilon^2)$ and $\Delta \Psi /\sH^2 =  \mathcal{O}(\epsilon)$ to leading order in Eqs.\,\eqref{pertexp}, where order $\epsilon^n$ is denoted by the sub- and superscript of $\Psi_n$ and $\omega_i^{(n)}$, respectively.
Plugging expansion \eqref{pertexp} into Eqs.\,\eqref{MasterEq} and demanding that the equations are fulfilled order by order, one can solve Eq.\,\eqref{MasterEq} iteratively.
At order $n$ only metric perturbations $\Psi_m$, $\omega^{(m)}_i$ with $m<n$ appear on the right hand side of Eq.\,\eqref{MasterEq}.  Therefore considering the system \eqref{MasterEq} at order $n$,  all the $\Psi_m$ and $\omega_i^{(m)}$ can be replaced by the lower order solutions obtained a step earlier.
From the solution for $\Psi$ and $\v{\omega}$ the density contrast $\delta$ and the velocity $v^i$ are obtained by perturbatively expanding the $00$ and $0i$ Einstein equation:
\begin{equation}
\delta = \frac{2}{3 \mathcal{H}^2} \Delta \Psi\,,\quad\quad
\v{v} =  -\frac{2}{3 \sH^2}\frac{ \v{\nabla}(\Psi'+ \mathcal{H}\Psi )- \tfrac{1}{4} \Delta \v{\omega}}{1 + \delta}\,.
\end{equation}

\subsection{First order}
We obtain neglecting the decaying mode
\begin{align}
&\Psi_1'' +3\mathcal H\Psi_1' = 0 \quad \Rightarrow \quad \Psi_1(\v{x},\tau)=\Psi_1(\v{x},\tau_{\rm ini}) =:\Psi_L(\v{x})\,.
\end{align}
There is no $\mathcal{O}(\epsilon)$ contribution to $\v{\omega}$. The solutions for $\delta_1$  and the peculiar velocity $\v{v}_1$ are 
\begin{align}
\delta_1 &= \frac{\tau^2}{6}\Delta\Psi_L =: \delta_L\,, \\
\v{v}_1 &= -\frac{2}{3\sH^2}\v{\nabla}(\Psi_1'+\sH \Psi_1)= -\frac{\tau}{3}\v{\nabla}\Psi_L\,,
\end{align}
which are the SPT results in first order.
\subsection{Second order} \label{sec:2order}
 To obtain the second order contribution we insert $\Psi_1$ into the right hand side of the $\Psi$-equation
\begin{align*}
\tag{\underline{3.6}}  \Psi_2''+3\sH\Psi_2' &= \frac{5}{3} \frac{\nabla_i\nabla_j}{\Delta} ({\Psi_L}_{,i}{\Psi_L}_{,j}) - \frac{1}{2} ({\Psi_L}_{,i})^2\,,  
\end{align*}
in which there is again no $\v{\omega}$, and separate the time dependence from the spatial one
\begin{align*}
\tag{\underline{3.11}} \Delta\Psi_2 &= \frac{\tau^2}{14} \left[ \frac{5}{3} (\Delta\Psi_L)^2 + \frac{7}{3} {\Delta\Psi_L}_{,i}  {\Psi_L}_{,i} + \frac{2}{3}  ({\Psi_L}_{,ij})^2 \right]\,. 
\end{align*}
The density contrast is related to the Newtonian potential
\begin{align*}
\delta_2 (\v{p}_3,\tau) &= -\frac{\tau^2}{6} p_3^2\Psi_2(\v{p}_3,\tau) =  \int \frac{\vol{3}{p_1}\vol{3}{p_2}}{(2\pi)^3} F_2(\v{p}_1,\v{p}_2) \delta_D(\v{p}_3-\v{p}_1-\v{p}_2)\delta_L(\v{p}_1,\tau)\delta_L(\v{p}_2,\tau)\,,\\
\tag{\underline{3.12}}  F_2(\v{p}_1,\v{p}_2)&= \frac{5}{7} + \frac{1}{2} (\v{p}_1\cdot\v{p}_2)\frac{p_1^2+p_2^2}{p_1^2p_2^2} +\frac{2}{7} \frac{(\v{p}_1\cdot\v{p}_2)^2}{p_1^2p_2^2} \,.
\end{align*}
where we obtained $F_2$ by reading off the coefficients and gradient structure from (\underline{3.11}), used $\delta=-\tau^2 k^2 \Psi/6$ and symmetrized the second term. The velocity is obtained from 
\begin{equation} \label{v2sol}
v_2^i= -\frac{2}{3 \mathcal{H}^2} (\Psi_2'+ \Psi_2 \mathcal{H})_{,i}  + \frac{4}{9 \mathcal{H}^4} (\Psi_1'+ \Psi_1 \mathcal{H})_{,i} \Delta \Psi_{1} + \frac{1}{6\mathcal{H}^2} \Delta \omega^{(2)}_i\,,
\end{equation}
with divergence $\theta = \v{\nabla} \cdot \v{v}$ given by
\begin{align}
-\theta_2 &= \delta_2'+\v{\nabla}\cdot(\delta_1 \v{v}_1)\,,
\end{align}
in which the $\Delta \omega^{(2)}_i$ drops out. Using the linear solution and $\delta_2$ we obtain its Fourier transform
\begin{align}
\notag-\frac{\theta_2(\v{p}_3,\tau)}{\sH} &= \mathcal{F}\left[2\delta_2 -\delta_L^2 -\v{\nabla}\delta_L\cdot\frac{\v{\nabla}\delta_L}{\Delta}\right](\v{p}_3)\\
 &= \int \frac{\vol{3}{p_1} \vol{3}{p_2}}{(2\pi)^3} G_2(\v{p}_1,\v{p}_2) \delta_D(\v{p}_3-\v{p}_1-\v{p}_2)\delta_L(\v{p}_1,\tau)\delta_L(\v{p}_2,\tau)\,,\\
\tag{\underline{3.15}} G_2(\v{p}_1,\v{p}_2) &= \frac{3}{7} + \frac{1}{2} (\v{p}_1\cdot\v{p}_2)\frac{p_1^2+p_2^2}{p_1^2p_2^2} +\frac{4}{7} \frac{(\v{p}_1\cdot\v{p}_2)^2}{p_1^2p_2^2} \,.
\end{align}
We see that in the derivation of $\delta_2$ and $\theta_2$ no $\v{\omega}$ contributed, such that one obtains the SPT result for $F_2$ and $G_2$ even in the case where one sets to zero vector perturbations. This can be also understood from the fluid-like system \eqref{Rjequations}, equivalent to the master equation \eqref{MasterEqPsi} with $\v{\omega}=0$ and neglecting \eqref{MasterEqOmeg}. The modified Euler equation \eqref{RjEuler} 
\begin{equation}
\nabla_i \left[ (1+\delta) \mathrm{Euler}^i \right] =0\,,\qquad \mathrm{Euler}^i := {v^i}'+\sH v^i+ v^i_{,j} v^j +\Psi_{,i}\,,
\end{equation}
at first order $\nabla_i \left(\mathrm{Euler}^i_1 \right) =0$ is simply the Bernoulli equation and therefore equivalent to \eqref{jEuler} in case of vanishing curl of $\v{v}$
\begin{equation}
 \mathrm{Euler}^i =0\quad \stackrel{\v{\nabla}\times \v{v}=0}{\Leftrightarrow}\quad   \nabla_i \mathrm{Euler}^i =0\,.
\end{equation}
At second order, the modified Euler equation \eqref{RjEuler}  takes the form 
\begin{equation} \label{v2RjEuler}
\nabla_i \left( \mathrm{Euler}^i_2 \right) = \nabla_i \left( \delta_1  \mathrm{Euler}^i_1\right)\,,
\end{equation}
which, upon using the 1st order solution, simplifies to
\begin{equation} \label{v2RjEulerSimp}
\nabla_i \left( \mathrm{Euler}^i_2 \right) = 0\,.
\end{equation}
Therefore $\theta_2$ is not modified, even though $v^i_2$ with $\v{\omega}=0$ is not curl-free and does not solve $\mathrm{Euler}^i_2 = 0$; it solves only \eqref{v2RjEulerSimp}. The $v^i_2$ including $\v{\omega}$,  \eqref{v2sol} does solve $\mathrm{Euler}^i_2 = 0$ and is curl-free.

To show this, let us calculate $\omega^{(2)}_i$ and the curl part of $v_2^i$.
At second order in perturbation theory we can neglect $\omega_i$ on the right hand side of \eqref{MasterEqOmeg} since they are always multiplied by $\Psi$ and therefore contribute to higher orders only. Using a growing mode ansatz for the time dependence we find $\omega^{(n)}_i(\v{x},\tau) = \tau^{(n-2)+1} \tilde{\omega}^{(n)}_i(\v{x})$ for $n\geq 2$, such that we obtain at second order 
\begin{equation} \label{solomega2}
\Delta \omega^{(2)}_i = \frac{8}{3 \mathcal{H}} \left(\frac{\nabla^i \nabla^m}{\Delta} - \delta^m_i\right) \nabla^k (\Psi_{1,k} \Psi_{1,m})\,.
\end{equation}
In Fourier space we can write down the kernel $\Omega_2^i$ for 
\begin{align}
\notag i\frac{\widetilde{\Delta \omega_i^{(2)}}(\v{p}_3,\tau)}{6 \sH^3} &= i \mathcal{F}\left[ \left(\frac{\nabla^i \nabla^m}{\Delta} - \delta^m_i\right) \nabla^k (\frac{\delta_{L,k}}{\Delta} \frac{\delta_{L,m}}{\Delta})\right](\v{p}_3)\\
 &= \int \frac{\vol{3}{p_1} \vol{3}{p_2}}{(2\pi)^3}  \Omega^i_2(\v{p}_1,\v{p}_2) \delta_D(\v{p}_3-\v{p}_1-\v{p}_2)\delta_L(\v{p}_1,\tau)\delta_L(\v{p}_2,\tau)\,,\\
\Omega^i_2(\v{p}_1,\v{p}_2) &= \left(\frac{(\v{p}_1+\v{p}_2)^i (\v{p}_1+\v{p}_2)^m}{(\v{p}_1+\v{p}_2)^2} - \delta ^m_i\right)(\v{p}_1+\v{p}_2)\cdot \frac{\v{p}_1}{p_1^2}\frac{ p_2^m}{{p_2^2}}\,.
\end{align}
The solution for the velocity $v_2^i$ is now curl-free, $(\delta^j_i-\nabla^i \nabla^j/\Delta ) v_2^j = 0$. To show this, we
expand the $0i$ equation
\begin{equation}
v^i = \frac{-2 (\Psi'+ \Psi \mathcal{H})_{,i} + \tfrac{1}{2} \Delta \omega_i}{3\mathcal{H}^2 +2 \Delta \Psi}
\end{equation}
to calculate the vector part of 
\begin{equation}
v_2^i= -\frac{2}{3 \mathcal{H}^2} (\Psi_2'+ \Psi_2 \mathcal{H})_{,i} + \frac{4}{9 \mathcal{H}^4} (\Psi_1'+ \Psi_1 \mathcal{H})_{,i} \Delta \Psi_{1}+ \frac{1}{6\mathcal{H}^2} \Delta \omega^{(2)}_i 
\end{equation}
using the transverse projector and the second order solution for $\omega_i^{(2)}$, Eq.\,\eqref{solomega2}. The simple calculation
\begin{align}
\left(\delta^j_i- \frac{\nabla^i \nabla^j}{\Delta} \right) v_2^j &= \frac{4}{9 \mathcal{H}^3} \left(\frac{\nabla^i \nabla^m}{\Delta} -\delta^m_i \right) (\Psi_{1,k} \Psi_{1,mk})\\
&= \frac{4}{9 \mathcal{H}^3} \left(\frac{\nabla^i \nabla^m}{\Delta} -\delta^m_i \right) \frac{1}{2} (\nabla \Psi)^2_{,m}
= 0
\end{align}
shows that $\v\nabla \times \v{v}_2=0$ and therefore $v_2^i$, including $\v{\omega}$, is a solution to $\mathrm{Euler}^i_2 =0$.
\subsection{Third order}\label{sec:thirdorder}
At third order in perturbation theory, $\v{\omega}$ will show up in $F_3$ and $G_3$. Therefore, contrary to claims in \citep{BR11}, perturbation theory based on $\v{\omega}=0$ is not equivalent to SPT. We will only consider the $\Psi$-equation, because at order $n=3$ only $\omega^{(2)}_i$ will be necessary to obtain $F_3$ and $G_3$. Taking into account that $\Psi_2' = \sH \Psi_2$ we obtain
\begin{align}
\tag{\underline{3.16} }\notag \Psi_3''+3\sH\Psi_3' &= - {\Psi_L}_{,i}{\Psi_2}_{,i}   + \frac{14}{3} \frac{\nabla_i\nabla_j}{\Delta} ({\Psi_L}_{,i}{\Psi_2}_{,j}) - \frac{4}{9\sH^2} \frac{\nabla_i\nabla_j}{\Delta}  \left( {\Psi_L}_{,i}{\Psi_L}_{,j}\Delta\Psi_L  \right)-\\
\quad&\quad \quad-\frac{1}{3 \sH}\frac{\nabla_i\nabla_j}{\Delta}\left( \Psi_{L,i} \Delta \omega_j^{(2)}\right)\,, \notag
\end{align}
where the second line is not present in Eq.\,3.16 of \cite{BR11}.
One can separate the time dependence from the spatial one and switch to Fourier space and use that a product in real space is a convolution in Fourier space. In addition we use that $\delta=-\tau^2 k^2 \Psi /6$ and insert the expression for $\Psi_2$ and $\omega_i^{(2)}$ to obtain
\begin{align}
\tag{\underline{3.20}} \delta_3 (\v{p}_4,\tau) &= -\frac{\tau^2}{6} p_4^2\Psi_3(\v{p}_4,\tau) \\
\notag &= \int \frac{\vol{3}{p_1}\vol{3}{p_2}\vol{3}{p_3}}{(2\pi)^6} F_3(\v{p}_1,\v{p}_2,\v{p}_3) \delta_D(\v{p}_4-\v{p}_1-\v{p}_2-\v{p}_3)\delta_L(\v{p}_1)\delta_L(\v{p}_2)\delta_L(\v{p}_3)\,.
\notag
\end{align}
The $F_3 = F_3^{(1)} +F_3^{(2)} +F_3^{(3)} +F_3^{(4)} $ can be decomposed into 4 parts of Eq.\,(\underline{3.16}), of which only the first 3 survive for $\v{\omega}=0$. In the following expressions $\v{p}_4=\v{p}_1+\v{p}_2+\v{p}_3$.
\begin{align}
\tag{\underline{3.20a}} F_3^{(1)} &= - \frac{1}{6} \frac{\v{p}_1\cdot(\v{p}_2+\v{p}_3)p_4^2}{p_1^2|\v{p}_2+\v{p}_3|^2 } F_2(\v{p}_2,\v{p}_3)\\\
\notag &= - \frac{1}{14} \frac{\v{p}_1\cdot(\v{p}_2+\v{p}_3)p_4^2}{p_1^2 p_2^2 p_3^2} \left\{\frac{5}{3} \frac{(\v{p}_2+\v{p}_3)\cdot \v{p}_2 \ (\v{p}_2+\v{p}_3)\cdot\v{p}_3}{|\v{p}_2+\v{p}_3|^2} -\frac{1}{2}(\v{p}_2\cdot\v{p}_3) \right\}\,, \\
\notag \\
\tag{\underline{3.20b}}F_3^{(2)} &= \frac{7}{9} \frac{(\v{p}_4\cdot\v{p}_1)[\v{p}_4\cdot(\v{p}_2+\v{p}_3)]}{p_1^2 |\v{p}_2+\v{p}_3|^2} F_2(\v{p}_2,\v{p}_3)\\
\notag  &= \frac{1}{3} \frac{(\v{p}_4\cdot\v{p}_1)[\v{p}_4\cdot(\v{p}_2+\v{p}_3)]}{p_1^2 p_2^2 p_3^2} \left\{ \frac{5}{3} \frac{(\v{p}_2+\v{p}_3)\cdot \v{p}_2 \ (\v{p}_2+\v{p}_3)\cdot\v{p}_3}{|\v{p}_2+\v{p}_3|^2} -\frac{1}{2}(\v{p}_2\cdot\v{p}_3)\right\}\,,\\
\notag \\
\tag{\underline{3.20c}} F_3^{(3)} &= -\frac{1}{9} \frac{(\v{p}_4\cdot\v{p}_1)(\v{p}_4\cdot\v{p}_2)p_3^2}{p_1^2 p_2^2 p_3^2}\,, \\
\tag{\underline{3.20d}} F_3^{(4)} &=- \frac{2}{9} \frac{\v{p}_4\cdot\v{p}_1}{p_1^2 p_2^2 p_3^2} p_4^j \left(\frac{(\v{p}_2+\v{p}_3)^j (\v{p}_2+\v{p}_3)^m}{|\v{p}_2+\v{p}_3|^2} - \delta^m_j\right) p_3^m (\v{p}_2+\v{p}_3)\cdot \v{p}_2\,.
\end{align}
In the main text we denoted the power spectrum calculated from $F^{\omega=0}_3 = F_3^{(1)} +F_3^{(2)} +F_3^{(3)}$ (the $F_3$ of \cite{BR11}) by $P^{\omega=0}$ . The power spectrum calculated from $F_3 = F_3^{(1)} +F_3^{(2)} +F_3^{(3)} +F_3^{(4)} $ was denoted by $P^{\rm SPT}$. Since the expression for $\delta_3$ only depends on the symmetrized part of $F_3$, denoted by $F_3^{\rm sym}$, one must use $F_3^{\rm sym}$ in the power spectrum. The  $F_3^{\rm sym}$ then has to be compared to the symmetrized versions from the literature in order to verify that our $F_3$ really coincides with SPT.  We did this in \textit{Mathematica} and checked against the formulas provided in \cite{GGRW86} and \cite{JB94}. We also explicitly show in this file the difference between the symmetrized $F^{\omega=0}_3$ and $F_3$. The \textit{Mathematica} notebook is enclosed in the arXiv source file for this document.

In the following we won't need the vector part of $\v{v}_3$, since we will only require $\v{\theta}_3$ for the velocity power spectrum at 1-loop order. We start with the $0j$-Einstein equation
\begin{equation}
v^i (1+\delta)= - \frac{2}{3\sH^2}\left(( \Psi'+ \Psi \mathcal{H})_{,i} - \tfrac{1}{4} \Delta \omega_i \right)
\end{equation}
expand to third order and take the divergence
%\begin{equation}
%\begin{split}
%v_i^{(3)}= -&\frac{2}{3 \mathcal{H}^2}\left(\Psi'_{3} +\mathcal{H}\Psi_{3}\right),_{i}+\frac{1}{6 \mathcal{H}^2}\Delta\omega^{(3)}_{i} \\
%+&\frac{4}{9 \mathcal{H}^4}\left(\Psi'_{2} +\mathcal{H}\Psi_{2}\right),_{i}\Delta\Psi_{1}+\frac{4}{9 \mathcal{H}^3}\Psi_{1,i}\Delta\Psi_{2}-\frac{1}{9 \mathcal{H}^4}\Delta\omega^{(2)}_{i}\Delta\Psi_1 \\
%-&\frac{8}{27\mathcal{H}^5}\left(\Delta\Psi_{1}\right)^2\Psi,_{i}
%\end{split}
%\end{equation}
\begin{equation}
-\theta_{3}=  3 \sH \delta_3+ \v{\nabla} \cdot (\v{v}_2 \delta_{1})+ \v{\nabla} \cdot (\v{v}_1 \delta_{2})\end{equation}
If $\delta_3$ and $\v{v}_2$ take their SPT forms also $\theta_3$ will coincide with SPT, see for instance Eq.\,(6b) of \cite{GGRW86}. Therefore in the case of dynamical $\v{\omega}$ we find $G_3 = G_3^{\rm SPT}$. In the case where we set $\v{\omega}=0$, the resulting expression differs due the differing $F_3^{\omega=0}$ and $\v{v}_2$ developing a nonzero curl
\begin{align}
\v{v}_2^{\omega=0} &= -2 \sH \frac{\v{\nabla} \delta_2}{\Delta} -\delta_1\v{v}_1\,.
\end{align}
Therefore $ G_3^{\omega=0}$ changes to
\begin{align}
\notag-\frac{\theta_3(\v{p}_4,\tau)}{\sH} 
 &= \int \frac{\vol{3}{p_1} \vol{3}{p_2}  \vol{3}{p_3}}{(2\pi)^6} G_3(\v{p}_1,\v{p}_2,\v{p}_3) \delta_D(\v{p}_4-\v{p}_1-\v{p}_2-\v{p}_3)\delta_L(\v{p}_1,\tau)\delta_L(\v{p}_2,\tau)\delta_L(\v{p}_3,\tau)\\
 G_3^{\omega=0}&= 3 F^{\omega=0}_3- \v{p}_4\cdot \left(\frac{\v{p}_{1}+\v{p}_{2}}{|\v{p}_{1}+\v{p}_{2}|^2} F_2(\v{p}_1,\v{p}_2) + \frac{\v{p}_2}{p^2_2}\right) - \v{p}_4\cdot \frac{\v{p}_1}{p_1^2} F_2(\v{p}_2,\v{p}_3)\,.
 \end{align}
  \begin{center}
\begin{figure}[t]
\includegraphics[width=0.49\textwidth]{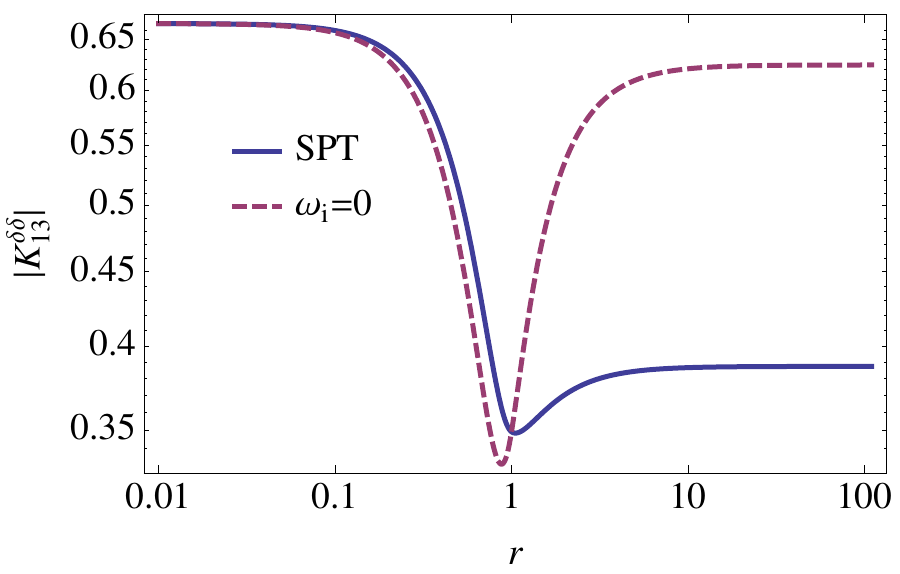}
\includegraphics[width=0.49\textwidth]{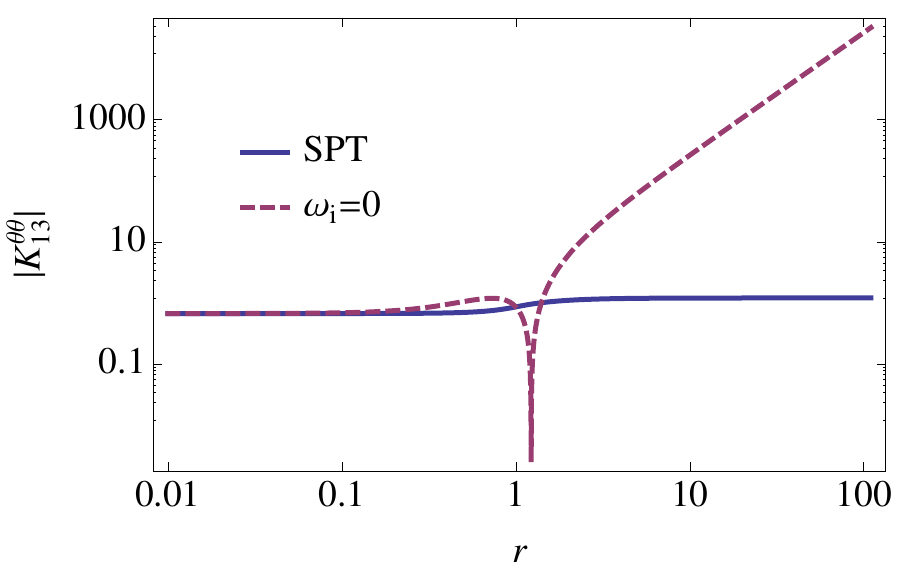}\\
\caption{\textit{Left}:  Comparison between integration kernels $K^{\delta \delta}_{13}$ of the 1-loop matter power spectrum for $P_{13, \delta\delta}$.   \textit{Right}: Comparison between the integration kernels $K^{\theta \theta}_{13}$ of the 1-loop velocity power spectrum kernel $P_{13,vv}$. For  $\omega=0$ the velocity kernel $K^{\theta \theta}_{13}$ goes like $r^2$ for $r \gg 1$ and leads to UV-divergent $P^{\omega=0}_{13,vv}$.}
\label{fig:CompRKernels}
\end{figure}
\end{center}
The velocity power spectrum is defined via $\langle \v{v}(\v{k})\cdot \v{v}(\v{p}) \rangle= (2\pi)^3 \delta_{\rm D}(\v{k} + \v{p})  P_{vv}(k)$. Because of the Dirac function the power spectrum splits into $P_{vv}(k)= P_{\theta \theta}(k) +P_{\mw \mw}(k)$. At 1-loop order using the symmetrized $G_2$, $G_3$ and $\Omega^i_2$ we find
\begin{subequations}
\label{Pvv}
\begin{align}
P_{vv}^{\omega=0}(k) &= \frac{1}{k^2}\left( \sH^2 P_{1}(k) +  P_{22,\theta\theta}(k)   +  P_{13,\theta\theta}^{\omega=0}(k) \right)+ P_{22,\mw\mw}^{\rm \omega =0}(k) \label{Pvvomega}\\
P_{vv}^{\text{SPT}}(k) &=   \frac{1}{k^2} \left(\sH^2P_{1}(k) +  P_{22,\theta\theta}(k) +  P_{13,\theta\theta}^{\rm SPT}(k)\right)\,,
\end{align}
\end{subequations}
with
\begin{subequations}
\begin{align}
P_{22,\mw\mw}^{\rm \omega =0}(k)  &= \frac{\sH^2 k}{2\cdot 4\pi^2} \int_0^\infty dr \int_{-1}^1 dx\ P_1(k\sqrt{1-2r x+r^2}) P_1(kr) \frac{(1-x^2)(1-2 rx)^2}{(1-2r x+r^2)^2}\\
&= P_{22,\omega \omega}^{\rm SPT}(k)\\
P_{13,\theta\theta}^{\omega=0}(k) &=   \frac{\sH^2 k^3}{42\cdot 4\pi^2} P_1(k) \int_0^\infty dr\ P_1(kr) \Bigg\{\frac{20}{ r^2}-\frac{319}{3}+\frac{76}{3}r^2-25r^4 +\label{Pthet13R}\\
\notag &\qquad\qquad\qquad\qquad\qquad\qquad\qquad+\frac{5}{2r^3}(r^2-1)^3(-5r^2+4)\ln\left|\frac{1+r}{1-r}\right| \Bigg\}\\
P_{13,\theta\theta}^{\rm SPT}(k) &=   \frac{\sH^2 k^3}{42\cdot 4\pi^2} P_1(k) \int_0^\infty dr  \ P_1(kr) \Bigg\{\frac{6}{ r^2}-41+2r^2-3r^4 +\label{Pthet13SPT}\\
\notag &\qquad\qquad\qquad\qquad\qquad\qquad\qquad+\frac{3}{2r^3}(r^2-1)^3(r^2+2)\ln\left|\frac{1+r}{1-r}\right| \Bigg\}\,. 
\end{align}
\end{subequations}
Here $P_{22,\theta\theta}$ is the standard expression, since $G_2$ does not depend on $\v{\omega}$, see (\underline{3.15}). The $P_{13,\theta\theta}$ are quite different, though. While $P_{13,\theta\theta}^{\rm SPT}$ converges,  $P^{\omega=0}_{13,\theta\theta}$ is UV-divergent for standard $P_1$, which go in the UV like $k^{0.96} (\ln k/k^2)^2 $. We plot in Fig.\,\ref{fig:CompRKernels} the integration kernels 
\begin{equation}
K^{\delta \delta}_{13}(r) = 6 r^2 \int_{-1}^{1} F^{\rm sym}_3(\v{q},-\v{q},\v{k}) dx\,,\qquad \qquad K^{\theta \theta}_{13}(r) = 6  r^2 \int_{-1}^{1} G^{\rm sym}_3(\v{q},-\v{q},\v{k}) dx\,,
\end{equation}
where $r = q/k$ and $x = \v{q}\cdot \v{k}/(q k) $. These kernels correspond to the curly brackets in  \eqref{P13R} and \eqref{P13SPT} as well as \eqref{Pthet13R} and \eqref{Pthet13SPT}, respectively.
 The  divergence of $P^{\omega=0}_{13,\theta\theta}$ is not cancelled by $P^{\omega=0}_{22,\mw \mw}$ in Eq.\,\eqref{Pvvomega}. The physical interpretation of this result remains unclear. Note that also for  power-laws $P_1\propto k^n$ UV-divergences can appear in SPT, whose physical meaning is not well understood, see for instance the discussion in \cite{BCGS02}, Sec 4.2.2. The 1-loop power spectra for  matter $P_{\delta \delta}$, \eqref{P13}, and momentum $P_{jj}$, \eqref{Pjj}, on the other hand are well behaved in the case of $\v{\omega}=0$, see also Fig.\,\ref{fig:CompR}. 

On the basis of these convergent expressions for $P_{\delta \delta}$ and $P_{jj}$ we can already  conclude that solutions obtained by setting $\v{\omega}=0$ and therefore solutions to \eqref{Rjequations} significantly deviate from solutions of the standard curl-free and pressureless fluid equations \eqref{jequations}. Having in addition a divergent  result for $P^{\omega=0}_{vv}$  might suggest the unphysical nature of \eqref{Rjequations}.
\section{Master Equation for perfect fluid and cosmological constant} \label{sec:masterwithpress}
We generalize the master equation to the case of an imperfect fluid with pressure/bulk viscosity $p$ and include a cosmological constant $\Lambda$. This serves as an example of how to generalize the master equation to arbitrary energy momentum tensors that extend the simple perfect and pressureless dust fluid $T_{\mu \nu} = \rho u_\mu u_\nu$. For instance, to model multi-streaming effects one might want to include an effective pressure and shear, see for instance \cite{BNSZ12}. If these additional tensors are functionals of $G_{00}$ and $G_{0i}$ only, they can be again eliminated from the Einstein equations. As an example we assume a perfect non-relativistic fluid with pressure $p = c_s^2 \bar{\rho}  \delta $. We simplify the energy momentum tensor according to the double expansion scheme and assuming non-relativistic velocities $\v{v}$.
\begin{subequations} \notag
\begin{align}
T_{\mu\nu} &=\rho u_\mu u_\nu + p (u_\mu u_\nu + g_{\mu\nu}) - \Lambda g_{\mu \nu}\,,\label{perfectEMT}\\
T_{00}&\simeq \rho u_0^2+ p (u_0^2 - a^2 ) - \Lambda g_{00} \simeq (\rho + \Lambda) a^2\\
T_{0i}&\simeq (\rho + p )u_0u_i\\
T_{ij}&\simeq(\rho + p )u_iu_j + (p-\Lambda)(1+2\Psi) a^2 \delta_{ij}
\end{align}
\end{subequations}
We can rewrite the first term in $T_{ij}$ using the Einstein equations
\begin{align} \notag
(\rho + p )u_iu_j  = &\frac{T_{0i} T_{0j}}{(\rho+p)u_0^2} = \frac{T_{0i} T_{0j}}{(\rho+\Lambda+p-\Lambda)u_0^2} = \frac{G_{0i} G_{0j}}{G_{00}+(p-\Lambda)a^2} \notag
\end{align}
The new source term therefore is
\begin{equation}\notag
\tilde{S}_{ij} = \frac{1}{2}(1+2\Psi)a^2 (p-\Lambda)\delta_{ij}+  \Psi_{,i}\Psi_{,j}+\frac{2}{3\mathcal{H}^2}\left\{  \frac{\left[\left(\Psi'+\mathcal{H}\Psi \right)_{,i}- \frac{1}{4}\Delta \omega_i\right] \left[\left(\Psi'+\mathcal{H}\Psi\right)_{,j}- \frac{1}{4}\Delta \omega_j\right]}{1+\frac{1}{3\mathcal{H}^2}\left[2\Delta\Psi+a^2(p-\Lambda)\right] }\right\}
\end{equation}
which should be used in \eqref{CoupledMasterEq} and \eqref{Rest}, with the right hand side of \eqref{CoupledMasterEqa}  replaced by
$$ \Psi'' +3\mathcal H\Psi' +\left(\mathcal{H}^2-2\frac{a''}{a}\right) \left(\frac{1}{2}-2\Psi\right)+\frac{1}{2}(\nabla\Psi)^2\,.$$
If the pressure is negligible compared to the energy density $p\ll \rho$, as it is the case for a fluid that models CDM, the only relevant new term is the first term in $\tilde{S}_{ij}$. In a similar fashion a shear viscosity $\eta$ as well as a bulk viscosity $\zeta$ could be added to the dust model $T_{\mu\nu}=\rho u_\mu u_\nu$, $$\Sigma_{\mu \nu} = -\eta  (u_\mu u^\alpha + \delta_\mu^\alpha)  (u^\beta u_\nu + \delta^\beta_\nu)(u_{(\alpha;\beta)}  -\frac{2}{3} g_{\alpha \beta} u^\rho_{;\rho}) - \zeta u^\rho_{;\rho} (u_\mu u_\nu + g_{\mu\nu})\,,$$ in which case the tensorial structure of the resulting $\bar{S}_{ij}$ would be more complicated. Assuming that in the small velocity limit, $\Sigma_{00} \ll T_{00}$ as well $\Sigma_{0i} \ll T_{0i}$ holds, the dominant parts of $\Sigma_{ij}$ are given by $\Sigma_{ij} \simeq - \eta a (v_{(i,j)}  -\frac{2}{3} \delta_{ij} \v{\nabla}\cdot \v{v})-\zeta \delta_{ij}\v{\nabla}\cdot \v{v}$. In this case any tensor depending only on $\delta$ and $\v{v}$, can be rewritten according to $00$ and $0i$ component of the Einstein equation
$$v_i = \frac{\left(\Psi'+\mathcal{H}\Psi \right)_{,i}- \frac{1}{4}\Delta \omega_i}{1+\frac{2}{3\mathcal{H}^2}\Delta\Psi}\,, \qquad \qquad \delta= \frac{2}{3\mathcal{H}^2}\Delta\Psi \,.$$
\end{document}